\documentclass[12pt]{iopart}
\usepackage{iopams}
\usepackage{graphicx}
\usepackage{cite}
\newcommand{\dif}{\mathrm{d}}
\newcommand{\erfc}{\mathop{\mathrm{erfc}}\nolimits}

\begin{document}
\title[Analytical approach to synchronous states of globally coupled noisy rotators]{Analytical approach to synchronous states of globally coupled noisy rotators}
\author{V~O~Munyaev$^1$, L~A~Smirnov$^{1, 2}$, V~A~Kostin$^{1, 2}$, G~V~Osipov$^1$ and A~Pikovsky$^{3, 1}$\footnote{Author to whom any correspondence should be addressed.}}
\address{$^1$ Department of Control Theory, Nizhny Novgorod State University, Gagarin Av.~23, Nizhny Novgorod, 603950, Russia.}
\address{$^2$ Institute of Applied Physics, Russian Academy of Sciences, Ul'yanova Str.~46, Nizhny Novgorod, 603950, Russia.}
\address{$^3$ Institute for Physics and Astronomy, University of Potsdam, Karl-Liebknecht-Str. 24/25, 14476 Potsdam-Golm, Germany.}
\ead{pikovsky@uni-potsdam.de}

\begin{abstract}
We study populations of globally coupled noisy rotators (oscillators with inertia) allowing a nonequilibrium transition from a desynchronized state to a synchronous one (with the non-vanishing order parameter).
The newly developed analytical approaches resulted in solutions describing the synchronous state with constant order parameter for weakly inertial rotators, including the case of zero inertia, when the model is reduced to the Kuramoto model of coupled noise oscillators.
These approaches provide also analytical criteria distinguishing supercritical and subcritical transitions to the desynchronized state.
All the obtained analytical results are confirmed by the numerical ones, both by direct simulations of the large ensembles and by the solution of the associated Fokker–-Planck equation.
We also propose generalizations of the developed approaches for setups where different rotators parameters (natural frequencies, masses, noise intensities, strengths and phase shifts in coupling) are dispersed.
\end{abstract}


\submitto{\NJP}

\maketitle

\section{Introduction}\label{sec:intro}

The Kuramoto model of globally coupled phase 
oscillators~\cite{Kuramoto-75} is a paradigmatic model to study synchronization 
phenomena~\cite{acebron_kuramoto_2005,gupta_kuramoto_2014,pikovsky_dynamics_2015,Gherardini_etal-18}.
In the thermodynamic limit, stationary (in a proper rotating reference frame)
synchronized states can be found analytically for an arbitrary distribution
of natural frequencies~\cite{Omelchenko-Wolfrum-12}. The idea is that for
any subgroup of oscillators having a certain natural frequency, a stationary distribution
in a constant mean field
and the corresponding subgroup order parameter
can be found analytically. Then the solution of the 
general self-consistent problem can be expressed via integrals of these subgroup order parameters.
Other analytical approaches for the Kuramoto model may be not restricted to stationary states, but usually
have other restrictions since they apply only for identical 
oscillators like the Watanabe--Strogatz theory~\cite{Watanabe-Strogatz-93}, 
or for a population with a Lorentzian distribution 
of natural frequencies like the Ott--Antonsen ansatz~\cite{Ott-Antonsen-08}.

An important generalization of the Kuramoto model considers
an ensemble of globally 
coupled rotators (that is, oscillators with inertia). 
It is in particular relevant for modelling power grid 
networks~\cite{filatrella_analysis_2008,grzybowski_synchronization_2016}.
Synchronization features of globally coupled rotators, both dterministic and noisy, 
have been widely 
studied~\cite{tanaka_first_1997,Acebron-Bonilla-Spigler-00,Olmi_etal-14,barre_bifurcations_2016,%
Gupta-Campa-Ruffo-18,Gao-Efstathiou-18};
in particular, an approach similar to the analytical description~\cite{Omelchenko-Wolfrum-12}
was developed~\cite{Komarov-Gupta-Pikovsky-14,gupta_kuramoto_2014,Campa-Gupta-Ruffo-15}.
However, for noisy coupled rotators, so far the stationary distributions 
were found only numerically. Similarly, there are no analytical formulae for the Kuramoto
model with noise.

In this paper, we present a fully analytical approach to the problem of
globally coupled noisy rotators for small intertia (small ``masses''). We analyse a 
stationary solution of the Fokker--Planck (Kramers) equation describing identical
noisy rotators driven by the mean field, 
and obtain a closed expression for the 
order parameter for this subgroup. 
Then, for an arbitrary distribution of natural frequencies,
a stationary solution for the global order parameter is expressed 
in an analytical parametric form. Based on this analysis, we derive the asymptotic form with respect to the small order parameter and use it to classify the transition to synchrony as a supercritical or a subcritical one.

The paper is organized as follows. We introduce the model of coupled noisy rotators 
in Section~\ref{sec:Model}, where we also formulate stationary equations for the distribution
density. A solution of these equation in the limit of small inertia terms is presented 
in Section~\ref{sec:MatrixMethod} (another derivation of this solution 
is given in Appendix~\ref{sec:CoordinateMethod}). Important particular 
cases of noise-free rotators and of noisy
coupled oscillators (i.e. the case of vanishing inertia term) are discussed in Section~\ref{sec:cases}.
There, we also give general expressions valid for small order parameters, i.e. close
to the transition point. Solutions for several popular distributions of the natural
frequencies (e.g., Gaussian, Lorentzian, etc.) with numerical examples
are presented in Section~\ref{sec:Examples}. In Section~\ref{sec:gen} we show how the
approach can be used for generic ensembles where not only natural frequencies, but 
other relevant parameters of rotators (masses, noise strengths, etc.) are distributed.
We conclude with Section~\ref{sec:con}.

\section{Noisy coupled rotators}\label{sec:Model}

In this paper we consider an ensemble of $N$ globally coupled rotators characterized by their angles $\varphi_n$ and velocities $\dot\varphi_n$ ($n = 1,2,\ldots,N$). 
The rotators are coupled via 
the complex mean field 
\begin{equation}
R\equiv re^{i\psi}=\frac{1}{N}\sum_{n=1}^N e^{i\varphi_n}
\label{eq:OrderParameter}
\end{equation}
and obey equations of motion
\begin{equation}
\eqalign{
\mu \ddot{\varphi}_n + \dot{\varphi}_n &= 
\omega_n + \frac{\varepsilon}{N} \sum_{n'=1}^{N} \sin\!\left(\varphi_{n'}-\varphi_n\right) + \sigma\xi_n\!\left(t\right) \cr
&=\omega_n + \varepsilon r \sin\!\left(\psi-\varphi_n\right) + \sigma\xi_n\!\left(t\right).}
\label{eq:InitialSystem}
\end{equation}
\looseness=-1 The unit of time is chosen so that the coefficient at the friction term ($\sim\dot\varphi_n$) 
is one and all the parameters are dimensionless (normalized by the friction coefficient). Parameter $\mu$ describes the 
mass of rotators (more precisely, $\mu$ should be called the moment of inertia of a rotator);
below, we focus on the overdamped case $\mu\ll 1$. Parameter $\varepsilon$ is the coupling strength. 
Parameters $\omega_n$ describe torques acting on rotators; 
we assume them to be distributed with a density $g(\omega)$. Note, that our approach can be
straightforwardly generalized to the case where other parameters are distributed as discussed in Section~\ref{sec:gen}.
Because uncoupled rotators have mean angular velocities $\omega_n$, we speak of ``natural frequencies''
instead of ``torques''. We do so also to make transparent a relation to the particular case $\mu=0$, where 
system~\eref{eq:InitialSystem} is nothing else but the standard Kuramoto model for globally coupled phase
oscillators with a distribution $g(\omega)$ of natural frequencies $\omega$. The rotators are acted upon by the independent white Gaussian noise forcings $\sigma\xi_n\left(t\right)$ with equal amplitudes $\sigma$, 
zero means $\langle\xi_n\!\left(t\right)\rangle = 0$, and 
auto-correlations $\langle\xi_n\!\left(t_1\right)\xi_{n'}\!\left(t_2\right)\rangle 
= 2\delta_{nn'}\delta\!\left(t_1-t_2\right)$ (where $\delta_{nn'}$ denotes the Kronecker delta, and $\delta\!\left(t\right)$ is the Dirac $\delta$-function). 
Whereas equations~\eref{eq:InitialSystem} are used for numerical simulations below, the 
analytical approach is developed in the thermodynamics limit $N\to\infty$.

In the mean-field coupling models, synchronization is one of the fundamental effects which is relevant to many systems. A transition to synchrony can be fully characterized 
in terms of the Kuramoto order parameter \eref{eq:OrderParameter}. Physically, the 
amplitude $r$ of this complex parameter describes the synchrony level of the elements 
belonging to the population considered. Nonvanishing $r$ indicates the collective synchronization. So, it is important to develop an universal analytical approach 
allowing to calculate $r$ value and predict the global evolution of an ensemble consisting 
of macroscopically large number of interacting units.

Note that model \eref{eq:InitialSystem} is not the most general one -- it might include a phase shift in coupling (which corresponds to the Kuramoto--Sakaguchi overdamped system).
We postpone the discussion of this case to Section~\ref{sec:gen}. Furthermore, for simplicity of presentation
we consider in Sections~\ref{sec:Model}--\ref{sec:Examples} only symmetric unimodal frequency distributions;
the discussion of asymmetric distributions is also postponed to Section~\ref{sec:gen}.

We now consider the limit of infinitely large number of elements, i.e., $N\to\infty$. 
Furthermore, we look for stationary synchronous states, i.e., those with constant
modulus of the order parameter and with a uniformly rotating angle:
$r=\mathrm{const}$, $\dot\psi=\Omega$. Such states are possible in the thermodynamics
limit only where the finite-size fluctuations vanish.
It is convenient to introduce a new angle variable related to the angle of the mean
field $\theta=\varphi-\psi$, $\dot\theta=u=\dot\varphi-\Omega$. Then, 
model \eref{eq:InitialSystem} can be formally rewritten as an infinite-dimensional system
\begin{equation}
\mu \ddot{\theta}_n + \dot{\theta}_n = 
\omega_n -\Omega- \varepsilon r \sin\!\left(\theta_n\right) + \sigma\xi_n\!\left(t\right),\quad
r=\langle e^{i\theta_{n}}\rangle.
\label{eq:InitialSystem1}
\end{equation}
Stochastic equation~\eref{eq:InitialSystem1} allows one to use the Fokker--Planck (Kramers)
equation for the probability density $P(\theta,u,t \mid \omega)$ for a subset of rotators having
natural frequency $\omega$ and to express the mean field as an integral over this density,
\begin{eqnarray*}
\partial_t P+\partial_\theta \left(u P\right)+\mu^{-1}\partial_u\left[ \left(-u+\omega-\Omega-\varepsilon r \sin\theta \right)P\right]=
\mu^{-2}\sigma^2\partial_u^2 P,\\
r=\int \dif\omega\, g(\omega)\int\dif\theta\, \int\dif u\,e^{i\theta} P(\theta,u,t \mid \omega).
\label{eq:fpe1}
\end{eqnarray*}
Because all the deterministic forces in this reference frame are constant, this density evolves toward
the stationary, time-independent, one. As the only parameters
governing this distribution are detuning $\nu=\omega-\Omega$ and forcing term $A=\varepsilon r$,
we seek this stationary distribution as a function depending on these parameters explicitly,
 $P_0(\theta,u \mid \nu,A)$. Then, the system takes the form
\begin{eqnarray}
u \partial_{\theta}P_0 = \mu^{-1}\partial_u\left(\left(u - \nu + A \sin \theta\right)P_0\right) +\mu^{-2}\sigma^2 \partial_u^2 P_0,
\label{eq:StationarySystem1}\\
r = \int\dif\nu\, g\!\left(\Omega+\nu\right) \int\dif\theta\, \int\dif u\,
	e^{i\theta} P_0\!\left(\theta, u \mid \nu,A\right), \quad A=\varepsilon r.
\label{eq:StationarySystem2}
\end{eqnarray}
The system \eref{eq:StationarySystem1} and \eref{eq:StationarySystem2} is in fact
a self-consistent system for determining the unknown order parameter $r$
(frequency $\Omega$ is obtained from the condition that $r$ is real).

\looseness=-1 Below, we restrict ourselves to symmetric distributions only, $g(\omega_0+\nu)=g(\omega_0-\nu)$.
Then, one may chose $\Omega=\omega_0$, which corresponds to the symmetric stationary density $P_0$, $P_0(\theta,u \mid \nu,A)=P_0(-\theta,-u \mid {-\nu},A)$. In this case, 
the integral in \eref{eq:StationarySystem2} is real,
which proofs that $\omega_0$ is a correct value of the frequency of the mean field $\Omega$. 
Note that there could also be other appropriate values of $\Omega$ besides $\omega_0$. 
These different values correspond to different synchronous branches.
Now, \eref{eq:StationarySystem2} provides a parametric solution
of the self-consistent problem since both $r$ and $\varepsilon$ are represented as
functions of a free parameter $A$,
\begin{equation*}
r(A)=\int{\dif\nu\, \dif\theta\,\dif u\, g\!\left(\omega_0+\nu\right) 
	e^{i\theta} P_0\!\left(\theta, u \mid \nu,A\right)}, \quad \varepsilon(A)= \frac{A}{r(A)}.
\end{equation*}
The only remaining act here is finding solution $P_0$ of~\eref{eq:StationarySystem1}.
In Refs.~\cite{Risken-89,Komarov-Gupta-Pikovsky-14}, a method
of matrix continuous fractions was employed to solve~\eref{eq:StationarySystem1} numerically,
another numerical approach is described in~\cite{Campa-Gupta-Ruffo-15}. 
In the next Section~\ref{sec:MatrixMethod}, we present an analytical solution for the overdamped case $\mu\ll 1$.

\section{Stationary distribution of the phases in the limit of small masses}\label{sec:MatrixMethod}

Here, we present one of the analytical methods for obtaining the stationary density $P_0$ from~\eref{eq:StationarySystem1} based on series representation. The second method is given in Appendix~\ref{sec:CoordinateMethod}.

\subsection{Matrix representation of the Fokker--Planck equation}\label{sec:MatrixRepresentation}

According to~\cite{Acebron-Bonilla-Spigler-00,Komarov-Gupta-Pikovsky-14}, we represent a 
stationary solution $P_0\!\left(\theta,u \mid \nu\right)$ of~\eref{eq:StationarySystem1} 
as a double series in the parabolic cylinder functions $\Phi_p\!\left(u\right)$ in $u$ and the
Fourier modes $e^{i q \theta}$ in $\theta$,
\begin{equation}
P_0\!\left(\theta,u \mid \nu,A\right) = \left(2 \pi\right)^{-1/2} \Phi_0\!\left(u\right) \sum_{p=0}^{+\infty} \sum_{q=-\infty}^{+\infty} 
a_{p q}\!\left(\nu,A\right) \Phi_p\!\left(u\right) e^{i q \theta}.
\label{eq:P1Series}
\end{equation}
Here functions $\Phi_p\!\left(u\right)$ are defined as
\begin{equation*}
\Phi_p\!\left(u\right) = \sqrt{\frac{\varkappa}{2^p p! \sqrt{\pi}}} \exp\left[-\varkappa^2 u^2/2\right] H_p\!\left(\varkappa u\right),
\label{eq:HermiteFunctions}
\end{equation*}
where $\varkappa = \sqrt{\mu / 2 \sigma^2}$ and $H_p\!\left(\varkappa u\right)$ are the Hermite polynomials.
Substituting expansion \eref{eq:P1Series} in~\eref{eq:StationarySystem1}, \eref{eq:StationarySystem2} and using 
orthogonality conditions for the basis functions, we obtain the infinite system for unknown coefficients
 $a_{pq}\!\left(\nu,A\right)$ where the order parameter $r$ is just proportional to one of the expansion coefficients,
\begin{equation}
\eqalign{\fl \sigma\sqrt{\frac{p + 1}{\mu}} i q a_{p+1,q} + \frac{p}{\mu} a_{p q} \\
-\sigma\sqrt{\frac{p}{\mu}}\left[\left(\frac{\nu}{\sigma^2}-i q\right) a_{p-1,q}
- i \frac{A}{2 \sigma^2} \left(a_{p-1,q+1} - a_{p-1, q-1}\right)\right) = 0,} \label{eq:SSM}
\end{equation}
\begin{equation}
\fl r = \sqrt{2 \pi}\int \dif\nu\, g\!\left(\omega_0+\nu\right) a_{0,1}^{*}\!\left(\nu,A\right).
\label{eq:lop}
\end{equation}
Thus, the main challenge is to find the quantity $a_{0,1}(\nu,A)$ via solving~\eref{eq:SSM}; this coefficient is nothing else
but the order parameter for the group of oscillators having natural frequency $\omega_0+\nu$; therefore, we call it
\emph{subgroup order parameter}. 
It is then used
 in~\eref{eq:lop} to find the total order parameter $r$ via averaging over $\nu$.
Below, we also use the symmetry property following from \eref{eq:SSM},
$a_{pq}\!\left(-\nu,A\right) = \left(-1\right)^p a_{pq}^{*}\!\left(\nu,A\right)$.

For the fixed parameters $A$ and $\nu$, the set of equations for $a_{pq}$ can be formally solved
by virtue of the matrix continuous fractions method~\cite{Risken-89,Komarov-Gupta-Pikovsky-14}.
According to this method, the expression for the subgroup order parameter 
$a_{0,1}\!\left(\nu,A\right)$ is as follows:
\begin{equation}
a_{0,1}\!\left(\nu,A\right) = \frac{S_{1,0}\!\left(\nu,A\right)}{\sqrt{2 \pi} S_{0,0}\!\left(\nu,A\right)},
\label{eq:a01General}
\end{equation}
where $S_{jk}\!\left(\nu,A\right)$ is the element of the infinite matrix $\mathbf{S}$, which is given by 
the recurrence formula
\begin{equation}
\mathbf{S} = \tilde{\mathbf{D}}^{-1}\left(\mathbf{I} - \mu \mathbf{D}\left(\mathbf{I} - \frac{\mu}{2} \mathbf{D}\left(\mathbf{I} - \frac{\mu}{3}\mathbf{D}\left(\mathbf{I} - \ldots\right)^{-1}\tilde{\mathbf{D}}\right)^{-1}\tilde{\mathbf{D}}\right)^{-1}\tilde{\mathbf{D}}\right).
\label{eq:MatrixH}
\end{equation}
where $\mathbf{I}$ is the identical matrix, and the infinite diagonal matrix $\mathbf{D}$ and the infinite tri-diagonal matrix $\tilde{\mathbf{D}}$ are defined as
\begin{equation}
D_{jk} = i k \delta_{j k}, \quad
\tilde{D}_{j k} = \left(i \sigma^2 k - \nu\right) \delta_{jk} + i \frac{A}{2} \left(\delta_{j,k - 1} - \delta_{j,k + 1}\right).
\label{eq:MatricesD}
\end{equation}

The main steps of the numerical procedure based on relations \eref{eq:a01General}--\eref{eq:MatricesD} employed for finding the order parameter with desired 
accuracy are described in detail in~\cite{Komarov-Gupta-Pikovsky-14}. Although this approach for 
the computation of the steady-state value of $r$ is efficient and has many potential
applications, there are several limitations in its use. In particular, this method has high time cost for the case of slowly descending distributions $g\left(\omega\right)$, e.g., 
for the Lorentz distribution. So, it is important to develop an analytical approximation for 
the calculation of the value $r$ in a nonequilibrium stationary state.

\subsection{Small mass approximation} 

Expression \eref{eq:MatrixH} allows for a perturbative expansion in the small parameter $\mu$.
In the first order in $\mu$, we obtain
\begin{equation}
\mathbf{S} = \tilde{\mathbf{D}}^{-1}\left(\mathbf{I} - \mu \mathbf{D}\tilde{\mathbf{D}}\right) + o\!\left(\mu\right).
\label{eq:MatrixH1Order}
\end{equation}
In order to evaluate the inverse matrix $\tilde{\mathbf{D}}^{-1}$, we denote the principal minors as
\begin{equation}
M_{jk} = \tilde{\mathbf{D}} \left[ \matrix{j & j + 1 & \ldots & k \cr j & j + 1 & \ldots & k} \right],
\label{eq:PrincipalMinors}
\end{equation}
and introduce a cutoff (cyclic) frequency $d\to+\infty$ for Fourier modes. Then, the elements of the inverse matrix $\tilde{\mathbf{D}}^{-1}$ take the form
\begin{eqnarray}
\tilde{D}_{jk}^{-1} &= \lim_{d \to +\infty} \left(i \frac{A}{2}\right)^{j-k} \frac{M_{-d,k-1} M_{j+1,d}}{\det \tilde{\mathbf{D}}} & \mbox{ for } j \ge k, \label{eq:InverseDElementsGeneral}\\
\tilde{D}_{jk}^{-1} &= (-1)^{j + k} \tilde{D}_{kj}^{-1} & \mbox{ for } j < k \nonumber.
\end{eqnarray}
According to the Laplace theorem, for all $j \ge k$ (for convenience, we take $M_{k+1,k} = 1$ and $M_{k+2,k} = 0$), the following representation holds
\begin{eqnarray}
\fl \det \tilde{\mathbf{D}} = M_{kj} M_{-d,k-1} M_{j+1,d} + \left(i \frac{A}{2}\right)^2 M_{k,j-1} M_{-d,k-1} M_{j+2,d} \nonumber\\
+ \left(i \frac{A}{2}\right)^2 M_{k+1,j} M_{-d,k-2} M_{j+1,d} + \left(i \frac{A}{2}\right)^4 M_{k+1,j-1} M_{-d,k-2} M_{j+2,d}.
\label{eq:LaplaceTheorem}
\end{eqnarray}

Substituting \eref{eq:LaplaceTheorem} in \eref{eq:InverseDElementsGeneral}, we obtain for $j \ge k$ that
\begin{eqnarray}
\fl {\tilde D}_{jk}^{-1} = \lim_{d \to +\infty}\left(i \frac{A}{2}\right)^{j-k} \left(M_{kj} + \left(i \frac{A}{2}\right)^2 M_{k+1,j} \frac{M_{-d,k-2}}{M_{-d,k-1}} \right. \nonumber\\
\left. + \left(i \frac{A}{2}\right)^2 M_{k,j-1} \frac{M_{j+2,d}}{M_{j+1,d}} + \left(i \frac{A}{2}\right)^4 M_{k+1,j-1} \frac{M_{-d,k-2}}{M_{-d,k-1}} \frac{M_{j+2,d}}{M_{j+1,d}}\right)^{-1}.
\label{eq:InverseDElements}
\end{eqnarray}

The principal minor $M_{jk}$ \eref{eq:PrincipalMinors} 
is found from a reccurent equation with fixed $j$, i.e., with fixed bottom right corner,
\begin{equation}
M_{kj} = \left(i \sigma^2 k -\nu\right) M_{k+1,j} + \left(i \frac{A}{2}\right)^2 M_{k+2,j}.
\label{eq:PrincipalMinorRecurrence1}
\end{equation}
Reccurent equation \eref{eq:PrincipalMinorRecurrence1} possesses a solution
\begin{eqnarray}
\fl M_{kj} 
\!=\! \left(i \frac{A}{2}\right)^{j-k+1}\! \frac{I_{-j-1-i\frac{\nu}{\sigma^2}}\!\left(\frac{A}{\sigma^2}\right) K_{-k+1-i\frac{\nu}{\sigma^2}}\!\left(-\frac{A}{\sigma^2}\right)
\!-\! K_{-j-1-i\frac{\nu}{\sigma^2}}\!\left(-\frac{A}{\sigma^2}\right) I_{-k+1-i\frac{\nu}{\sigma^2}}\!\left(\frac{A}{\sigma^2}\right)}{I_{-j-1-i\frac{\nu}{\sigma^2}}\!\left(\frac{A}{\sigma^2}\right) K_{-j-i\frac{\nu}{\sigma^2}}\!\left(-\frac{A}{\sigma^2}\right)
\!-\! K_{-j-1-i\frac{\nu}{\sigma^2}}\!\left(-\frac{A}{\sigma^2}\right) I_{-j-i\frac{\nu}{\sigma^2}}\!\left(\frac{A}{\sigma^2}\right)}.
\label{eq:PrincipalMinorSolution1}
\end{eqnarray}
Here $I_z$ and $K_z$ denote the modified Bessel functions of the first and second kind, respectively, of the order $z$. Using properties of the modified Bessel functions, we evaluate the limit
\begin{equation}
\lim_{d \to +\infty}\frac{I_{d -i\frac{\nu}{\sigma^2}}\!\left(\frac{A}{\sigma^2}\right)}{K_{d-i\frac{\nu}{\sigma^2}}\!\left(-\frac{A}{\sigma^2}\right)} = 0.
\label{eq:BesselIKLimit}
\end{equation}
Employing both \eref{eq:PrincipalMinorSolution1} and \eref{eq:BesselIKLimit}, we obtain
\begin{equation}
\lim_{d \to +\infty}\frac{M_{-d,k-2}}{M_{-d,k-1}} = - \left(i \frac{A}{2}\right)^{-1} \frac{I_{-k+1-i\frac{\nu}{\sigma^2}}\!\left(\frac{A}{\sigma^2}\right)}{I_{-k-i\frac{\nu}{\sigma^2}}\!\left(\frac{A}{\sigma^2}\right)}.
\label{eq:lLimit}
\end{equation}

Similarly, we find a representation of the principal minor $M_{kj}$, using its
expansion at fixed $k$, i.e., at fixed upper left corner,
\begin{equation*}
M_{kj} = \left(i \sigma^2 j - \nu\right) M_{k,j-1} + \left(i \frac{A}{2}\right)^2 M_{k,j-2},
\label{eq:PrincipalMinorRecurrence2}
\end{equation*}
which yields
\begin{equation}
\fl M_{kj} = \left(i \frac{A}{2}\right)^{j-k+1} \frac{I_{k-1+i\frac{\nu}{\sigma^2}}\!\left(-\frac{A}{\sigma^2}\right) K_{j+1+i\frac{\nu}{\sigma^2}}\!\left(\frac{A}{\sigma^2}\right)-K_{k-1+i\frac{\nu}{\sigma^2}}\!\left(\frac{A}{\sigma^2}\right) I_{j+1+i\frac{\nu}{\sigma^2}}\!\left(-\frac{A}{\sigma^2}\right)}{I_{k-1+i\frac{\nu}{\sigma^2}}\!\left(-\frac{A}{\sigma^2}\right) K_{k+i\frac{\nu}{\sigma^2}}\!\left(\frac{A}{\sigma^2}\right)-K_{k-1+i\frac{\nu}{\sigma^2}}\!\left(\frac{A}{\sigma^2}\right) I_{k+i\frac{\nu}{\sigma^2}}\!\left(-\frac{A}{\sigma^2}\right)}.
\label{eq:PrincipalMinorSolution2}
\end{equation}
Using expressions \eref{eq:PrincipalMinorSolution2} and \eref{eq:BesselIKLimit}, we find
\begin{equation}
\lim_{d \to +\infty}\frac{M_{j+2,d}}{M_{j+1,d}} = - \left(i \frac{A}{2}\right)^{-1} 
\frac{I_{j+1+i\frac{\nu}{\sigma^2}}\!\left(-\frac{A}{\sigma^2}\right)}{I_{j+i\frac{\nu}{\sigma^2}}\!\left(-\frac{A}{\sigma^2}\right)}.
\label{eq:rLimit}
\end{equation}

Combination of expressions \eref{eq:MatricesD}, \eref{eq:MatrixH1Order}, 
\eref{eq:PrincipalMinors}, \eref{eq:InverseDElements}, \eref{eq:lLimit}, and \eref{eq:rLimit} 
with the general formula \eref{eq:a01General} results in the closed-form formula to the first order in 
the mass parameter\,$\mu$,
\begin{equation}
a_{0,1}\!\left(\nu,A\right) = \frac{1}{\sqrt{2 \pi}} \frac{I_{1+i\frac{\nu}{\sigma^2}}\!\left(\frac{A}{\sigma^2}\right)}
{I_{i\frac{\nu}{\sigma^2}}\!\left(\frac{A}{\sigma^2}\right)}
\left(1-\mu \frac{\sigma^2}{\pi}\frac{\sin\!\left(i \pi \frac{\nu}{\sigma^2}\right)}
{I_{-i \frac{\nu}{\sigma^2}}\!\left(\frac{A}{\sigma^2}\right)I_{i \frac{\nu}{\sigma^2}}
\!\left(\frac{A}{\sigma^2}\right)}\right) + o\!\left(\mu\right).
\label{eq:a01Solution}
\end{equation}

An alternative derivation of our main result \eref{eq:a01Solution} is based on the method of moments for the density and on elimination of the velocity; it is presented in~\ref{sec:CoordinateMethod}.

\section{Limiting cases}\label{sec:LimitingCases}\label{sec:cases}

Above, we have derived a general expression for the subgroup order parameter~\eref{eq:a01Solution}. Here, we discuss its form in several important particular cases.

\subsection{Noise-free case}\label{sec:CaseWithoutNoise}

For purely deterministic rotators, we have to set $\sigma \to 0$. To find the noise-free limit of general expression \eref{eq:a01Solution}, it is convenient to rewrite it in an equivalent form
\begin{equation}
\fl a_{0,1}\!\left(\nu,A\right) = \frac{1}{\sqrt{2 \pi}} \frac{I_{1+i\frac{\nu}{\sigma^2}}\!\left(\frac{A}{\sigma^2}\right)}
{I_{i\frac{\nu}{\sigma^2}}\!\left(\frac{A}{\sigma^2}\right)}
\left(1-\mu \frac{A}{2}\left(\frac{I_{1+i\frac{\nu}{\sigma^2}}\!\left(\frac{A}{\sigma^2}\right)}
{I_{i\frac{\nu}{\sigma^2}}\!\left(\frac{A}{\sigma^2}\right)} - \frac{I_{-1-i\frac{\nu}{\sigma^2}}\!\left(\frac{A}{\sigma^2}\right)}
{I_{-i\frac{\nu}{\sigma^2}}\!\left(\frac{A}{\sigma^2}\right)}\right)\right) + o\!\left(\mu\right).
\label{eq:a01SolutionEquivalent}
\end{equation}
So, it is neccesary to calculate limits at $\sigma\to 0$ of the two fractions of the modified Bessel functions in \eref{eq:a01SolutionEquivalent}.
After evaluating the limits, one obtains
\begin{eqnarray}
\lim_{\sigma \to 0} \frac{I_{1 + i\frac{\nu}{\sigma^2}}\!\left(\frac{A}{\sigma^2}\right)}{I_{i\frac{\nu}{\sigma^2}}\!\left(\frac{A}{\sigma^2}\right)} &= \cases{-i\frac{\nu}A + \sqrt{1 - \frac{\nu^2}{A^2}}, & $\nu > -A$ \nonumber\\
-i\frac{\nu}A - \sqrt{1 - \frac{\nu^2}{A^2}}, & $\nu \le -A$,} \\
\lim_{\sigma \to 0} \frac{I_{- 1 - i\frac{\nu}{\sigma^2}}\!\left(\frac{A}{\sigma^2}\right)}{I_{-i\frac{\nu}{\sigma^2}}\!\left(\frac{A}{\sigma^2}\right)} &= \cases{-i\frac{\nu}A + \sqrt{1 - \frac{\nu^2}{A^2}}, & $\nu \le A$ \nonumber\\
-i\frac{\nu}A - \sqrt{1 - \frac{\nu^2}{A^2}}, & $\nu > A$,}
\end{eqnarray}
The resulting expression for the subgroup order parameter (in the first order in $\mu$) is
\begin{equation*}
\fl a_{0,1}\!\left(\nu,A\right) = o\!\left(\mu\right) + \frac{1}{\sqrt{2\pi}A} \cases{
i\left(-\nu-\sqrt{\nu^2-A^2}\right)\left(1+i\mu \sqrt{\nu^2-A^2}\right), & $\nu<-A$,\\
\left(-i\nu+\sqrt{A^2-\nu^2}\right), & $\left|\nu\right| \le A$,\\
i\left(-\nu+\sqrt{\nu^2-A^2}\right)\left(1-i\mu \sqrt{\nu^2-A^2}\right), & $\nu >A$.}
\label{eq:a01SolutionSigma0}
\end{equation*}

\subsection{Massless rotators: Kuramoto model}\label{sec:CaseWithoutInertia}

In the massless case ($\mu=0$), the problem~\eref{eq:InitialSystem} 
is in fact the Kuramoto model with noise. The analytical expression 
of the local order parameter is exact in the thermodynamics limit (we can now omit the first index) as follows:
\begin{equation}
a_{1}\!\left(\nu,A\right) = \frac{1}{\sqrt{2 \pi}} \frac{I_{1+i\frac{\nu}{\sigma^2}}\!\left(\frac{A}{\sigma^2}\right)}
{I_{i\frac{\nu}{\sigma^2}}\!\left(\frac{A}{\sigma^2}\right)}.
\label{eq:klop}
\end{equation}
Correspondingly, the expression for the full order parameter is also exact,
\begin{equation}
r = \int \dif\nu\, g\!\left(\omega_0+\nu\right) \frac{I_{1-i\frac{\nu}{\sigma^2}}\!\left(\frac{A}{\sigma^2}\right)}
{I_{-i\frac{\nu}{\sigma^2}}\!\left(\frac{A}{\sigma^2}\right)}.\label{eq:kurop}
\end{equation}
The subgroup order function~\eref{eq:klop} has no poles in lower half-plane, and the integral
in \eref{eq:kurop} can be evaluated for suitable distributions $g(\omega)$ via residues. For example, for the
Lorentz distribution
\begin{equation}
g\!\left(\omega_0+\nu\right) = \frac{\gamma}{\pi\left(\nu^2+\gamma^2\right)},
\label{eq:LorentzDistribution}
\end{equation}
the resulting expression for $r$ reads
\begin{equation}
r=\frac{I_{1+\frac{\gamma}{\sigma^2}}\!\left(\frac{A}{\sigma^2}\right)}
{I_{\frac{\gamma}{\sigma^2}}\!\left(\frac{A}{\sigma^2}\right)}.
\label{eq:rWithoutInertia}
\end{equation}
A more elaborated application of the theory to the Kuramoto model with noise will
be presented elsewhere.

\subsection{Synchronization transition}\label{sec:GeneralAnalysis}

Here, we employ the general parametric representation of the order parameter as a function of the coupling constant,
\begin{equation}
r = \sqrt{2 \pi}\int \dif\nu\, g\!\left(\omega_0+\nu\right) a_{0,1}^{*}\!\left(\nu,A\right),\quad \varepsilon=\frac{A}{r}
\label{eq:gpr}
\end{equation}
with $a_{0,1}\!\left(\nu,A\right)$ given by \eref{eq:a01Solution} to characterize the synchronization transition, i.e.,
to characterize states with the order parameter close to zero. All formulas below are valid
in the first order in small mass $\mu$ like \eref{eq:a01Solution} and are therefore approximate, but for the sake of simplicity we write them as exact relations below.

We expand $r(A)$ in the Taylor series for small values of $A$. This expansion
contains only odd powers of $A$ since \eref{eq:a01Solution} is an odd function of $A$,
\begin{equation}
r\!\left(A\right)=C_0 A + C_1 A^3 + C_2 A^5 + \ldots.
\label{eq:rExpansion}
\end{equation}
The three initial coefficients in expansion~\eref{eq:rExpansion} are as follows:
\begin{eqnarray}
C_0 &= \frac{1}{2} \int\dif y\, \frac{g\!\left(\omega_0+\sigma^2 y\right)}{1+y^2}\left(1-\mu\sigma^2 y^2\right), \nonumber \\
C_1 &= - \frac{1}{8 \sigma^4} \int\dif y\, \frac{g\!\left(\omega_0+\sigma^2y\right)}{\left(1+y^2\right)^2\left(4+y^2\right)}\left(2 \left(1-2 y^2\right) - \mu\sigma^2 y^2 \left(13 + y^2\right)\right), \label{eq:rC012}
\end{eqnarray}
\begin{eqnarray}
\fl C_2 = \frac{1}{16 \sigma^8} \nonumber\\
\fl {}\times\int\dif y\,\frac{g\!\left(\omega_0+\sigma^2 y\right)}{\left(1+y^2\right)^3\left(4+y^2\right)\left(9+y^2\right)}\left(2\left(3-17 y^2+4 y^4\right)+\mu\sigma^2 y^2\left(-113+32 y^2+y^4\right)\right). \nonumber 
\end{eqnarray}
In the massless case $\mu=0$, the expressions for $C_0,C_1$ have been obtained in~\cite{Bonilla-Neu-Spigler-92}.

The nontrivial branch of solutions $r(\varepsilon)$ starts at 
\begin{equation*}
\varepsilon_c^{\left(1\right)} = 1/C_0.
\label{eq:Kc1}
\end{equation*}
In the noiseless limit, $\sigma\to 0$, the critical value of the coupling parameter
can be expressed as
\begin{equation*}
\left.\varepsilon_c^{\left(1\right)}\right|_{\sigma=0} = \frac{2}{\pi g\!\left(\omega_0\right)-\mu}.
\label{eq:Kc1-1}
\end{equation*}
This expression coinsides, in the first order in $\mu$, with the result for deterministic
rotators obtained in~\cite{Gao-Efstathiou-18}.

It is instructive to compare this critical value with the expression for the stability loss of the asynchronous
state $r=0$ (equation~(24) in~\cite{Acebron-Bonilla-Spigler-00}). In our notations, this general formula 
for the imaginary part of the eigenvalue $x$, valid also
for large masses $\mu$, reads
\begin{equation}
1=\frac{\mu \hat\varepsilon e^{\mu\sigma^2}}{2}\sum_{p=0}^\infty \frac{(-\mu\sigma^2)^2 \left(1+\frac{p}{\mu\sigma^2}
\right)}{p!}\int_{-\infty}^\infty \dif\omega\, \frac{g(\omega)}{\mu\sigma^2+p+i(\mu\omega+x)}.
\label{eq:stabA}
\end{equation}
For small $\mu$, this experssion can be simplified. Assuming $x=x_0+\mu x_1+\mu^2 x_2$
and substituting this in \eref{eq:stabA}, we find $x_0=x_2=0$.
For a unimodal frequency distribution symmetric around $\omega_0$, a solution $x_1=-\omega_0$
exists, which yields 
\begin{equation*}
\hat\varepsilon^{-1}=\frac{\sigma^2}{2}\int_{-\infty}^\infty \dif\omega\,\frac{g(\omega)}{\sigma^4+(\omega-\omega_0)^2}
-\frac{\mu}{2}\int_{-\infty}^\infty \dif\omega\,\frac{g(\omega)(\omega-\omega_0)^2}{\sigma^4+(\omega-\omega_0)^2}+o(\mu).
\label{eq:epscr}
\end{equation*}
One can easily see that $\hat\varepsilon=\varepsilon_c^{\left(1\right)}$, which means that the branch of stationary
solutions joins the axis $r=0$ in the $\left(r,\varepsilon\right)$ plane exactly where the instability of the asynchronous state first occurs.

Depending on the sign of the coefficient $C_1$, there are two possibilities:
\paragraph{Supercritical transition} occurs for $C_1<0$. Here one observes a continuous (second-order) transition with the solution branch
\begin{equation*}
r = C_0^2 \sqrt{\left(\varepsilon_c^{\left(1\right)} - \varepsilon\right)/C_1} 
\label{eq:rKMainAsymptotic}
\end{equation*}
existing for $\varepsilon>\varepsilon_c^{\left(1\right)}$. 
\paragraph{Subcritical transition} occurs for $C_1>0$. Here, the branch of solutions exists for $\varepsilon<\varepsilon_c^{\left(1\right)}$. 
If $C_2<0$, one can estimate that this branch spreads till the minimal value at 
\begin{equation*}
\varepsilon_c^{\left(2\right)} \approx \left(C_0 - \frac{C_1^2}{4 C_2}\right)^{-1}.
\label{eq:Kc2}
\end{equation*}
Here the synchronization transition is discontinuous (a first-order transition).

We see, that the type of the transition is determined by the sign of $C_1$. Because
this coeffictient depends on the mass $\mu$, there is a critical value of the mass at
which the type of the transition changes,
\begin{equation}
\fl \mu^{*} = \frac{2}{\sigma^2} \left(\int\dif y \,\frac{g\!\left(\omega_0+\sigma^2 y\right)\left(1-2 y^2\right)}{\left(1+y^2\right)^2\left(4+y^2\right)}\right)\left(\int\dif y \,\frac{g\!\left(\omega_0+\sigma^2 y\right)y^2\left(13+y^2\right)}{\left(1+y^2\right)^2\left(4+y^2\right)}\right)^{-1}.
\label{eq:mucr}
\end{equation}
We stress that this expression is valid only if the value of $\mu^*$ is numerically small: $\mu^*\ll 1$. 

\section{Examples}\label{sec:Examples}
Here we present explicit calculations of the synchronization transition for several commonly explored
distributions of the frequencies. Each of these distributions is characterized by a width $\gamma$, and in all cases the result depends on the dimensionless parameter $\xi=\gamma/\sigma^2$, which measures
relative influence of the frequency dispersion and noise on the transition.

\subsection{Gaussian distribution}\label{sec:GaussDistribution}

\begin{figure}[h]
	\centering
	\includegraphics[width=9.675cm]{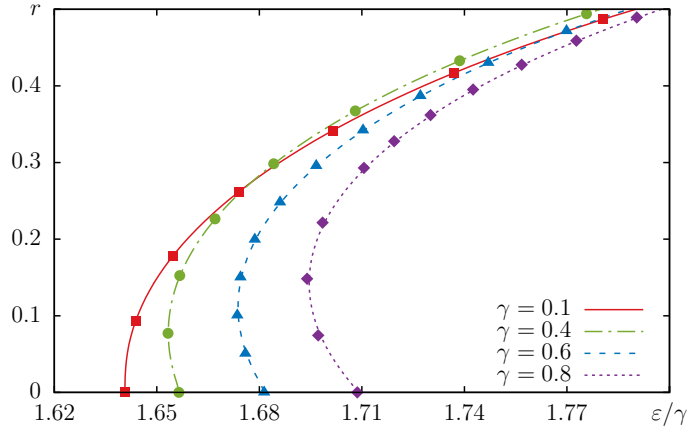}
\caption{
Branches of the desynchronized stationary solutions $r\!\left(\varepsilon\right)$ for the Gaussian distribution of frequencies
with different widths $\gamma$: the stationary
order parameter $r$ vs normalized coupling $\varepsilon/\gamma$ is shown.
The other parameters are $\mu=0.1$, $\sigma = 0.05$.
Curves are obtained via analytical solution \eref{eq:a01Solution} in the first order in $\mu$;
markers are obtained through numerical solutions of~\eref{eq:a01General}.
Small deviations of the analytical solution from the numerical one are noticeable for large
values of $\gamma$ only.
}
	\label{fig_r_Gauss}
\end{figure}

The Gaussian distribution of the rotators' frequencies is
\begin{equation*}
g\!\left(\omega\right) = \frac{1}{\sqrt{2\pi \gamma^2}} e^{-\frac{\left(\omega-\omega_0\right)^2}{2 \gamma^2}}.
\label{eq:GaussianDistribution}
\end{equation*}
For this distribution, the coefficients are
\begin{eqnarray*}
C_0 = \frac{1}{\sigma^2}\left(\sqrt{\frac{\pi}{8\xi^2}}e^{\frac{1}{2\xi^2}}\erfc\!\left(\frac{1}{\sqrt{2}\xi}\right) \left(1 + \mu\sigma^2\right) - \frac{\mu\sigma^2}{2}\right), \\
\fl \mu^{*} = \frac{1}{\sigma^2}\frac{2\pi e^{\frac{1}{2\xi^2}}\left(\xi^2-1\right)\erfc\!\left(\frac{1}{\sqrt{2}\xi}\right) + 2\xi\left(\sqrt{2\pi}\left(1-2\xi^2\right) + 3\pi e^{\frac{2}{\xi^2}}\xi\erfc\!\left(\frac{\sqrt{2}}{\xi}\right)\right)}{\xi^2\left(\sqrt{2\pi}\xi\left(10 + \xi^2\right) - 4\pi\left( e^{\frac{1}{2\xi^2}}\erfc\!\left(\frac{1}{\sqrt{2}\xi}\right) + 3e^{\frac{2}{\xi^2}}\erfc\!\left(\frac{\sqrt{2}}{\xi}\right)\right)\right)},
\label{eq:KcGaussianDistribution}
\end{eqnarray*}
where $\erfc$ is the complementary error function. We illustrate several cases with supercritical and subcritical transition 
in Figure~\ref{fig_r_Gauss}. Here parameter $\mu$ is not too small, nevertheless
the correspondence between the analytical and numerical results is very good. One can see that 
for a narrow distribution ($\gamma=0.1$), the synchronization transition is supercritical (solid curve),
whereas for the broader distributions, it is subcritical. 

\subsection{Lorentz distribution}\label{sec:UniformDistribution}

\begin{figure}[h]
	\centering
	\includegraphics[width=9.675cm]{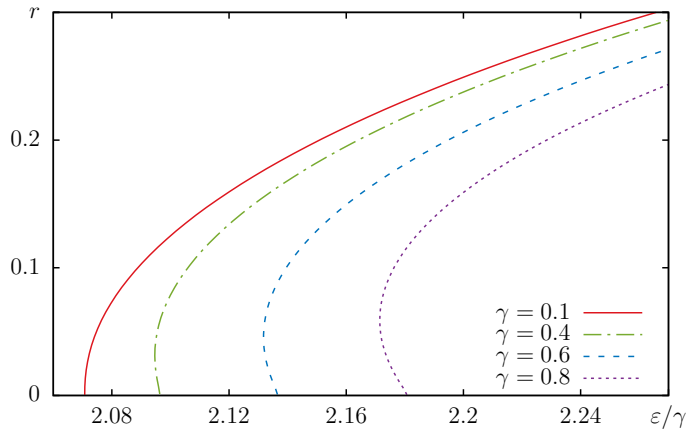}
	\caption{Branches of stationary solutions $r(\varepsilon)$ for
$\mu=0.1$, $\sigma = 0.05$, and different $\gamma$ for the Lorentz distribution of frequencies. 
Only the analytical solutions obtained from~\eref{eq:a01Solution} are presented, because precise 
numerical solution of the full problem is hard due to the broad tails of the distribution.}
	\label{fig_r_Lorenz}
\end{figure}

In the case of the Lorentz distribution \eref{eq:LorentzDistribution}, the expressions for 
characterization of the synchronization transition are
as follows:
\begin{equation*}
\fl C_0 = \frac{1}{2 \sigma^2}\frac{1-\mu\sigma^2\xi}{1+\xi}, \quad
C_1 = \frac{1}{8\sigma^6} \frac{-1+\mu\sigma^2\xi\left(3+\xi\right)}{\left(1+\xi\right)^2\left(2+\xi\right)}, \quad
\mu^{*} = \frac{1}{\sigma^2} \frac{1}{\xi\left(\xi+3\right)}.
\label{eq:C01mucrCauchyDistribution}
\end{equation*}
Different cases of supercritical and subcritical transitions are 
illustrated in Figure~\ref{fig_r_Lorenz}. The results are very close to the results shown in Figure~\ref{fig_r_Gauss} for the Gaussian distribution. The transition is supercritical for a narrow distributions
of natural frequencies and subcritical for larger values of $\gamma$.

\subsection{Bimodal distribution}\label{sec:BiharmonicDistribution}

For a bimodal distribution
\begin{equation*}
g\!\left(\omega\right) = \frac{1}{2}\delta\!\left(\omega-\omega_0-\gamma\right)+\frac{1}{2}\delta\!\left(\omega-\omega_0+\gamma\right),
\label{eq:BiharmonicDistribution}
\end{equation*}
we find
\begin{equation*}
\fl C_0 = \frac{1}{2\sigma^2}\frac{1-\mu\sigma^2\xi^2}{1+\xi^2}, \quad
C_1 = \frac{1}{8 \sigma^6}
\frac{\left(4+13 z\right)\xi^2-2+\mu\sigma^2\xi^4}{\left(1+\xi^2\right)^2\left(4+\xi^2\right)}, \quad
\mu^{*} = \frac{2}{\sigma^2}\frac{1-2\xi^2}{\xi^2\left(13+\xi^2\right)}.
\label{eq:rBiharmonicDistribution}
\end{equation*}
Furthermore, here the integral in \eref{eq:gpr} can be evaluated in terms of the modified Bessel functions,
\begin{equation*}
r = \Re\!\left( \frac{I_{1+i\xi}\!\left(\frac{A}{\sigma^2}\right)}{I_{i\xi}\!\left(\frac{A}{\sigma^2}\right)}\left(1-\mu\frac{\sigma^2}{\pi}\frac{\sin\!\left(i \pi \xi\right)}{I_{-i \xi}\!\left(\frac{A}{\sigma^2}\right)I_{i \xi}\!\left(\frac{A}{\sigma^2}\right)}\right)\right) + o\!\left(\mu\right),
\label{eq:rBiharmonicDistribution1}
\end{equation*}
Remarkably, for $\gamma>\sigma^2/\sqrt{2}$ we have $\mu^*<0$, which means that in this range of parameters
the stationary branch bifurcates subcritically for any masses.

Noteworthy, for the bimodal distribution, non-stationary (time-periodic) solutions can dominate the transition,
as have been demonstrated in~\cite{Bonilla-Neu-Spigler-92,Acebron-Bonilla-Spigler-00}, and the above-presented analysis of stationary states solves only a part of the problem.

\subsection{Uniform distribution}\label{sec:LorentzDistribution}

Here we consider a uniform distribution
\begin{equation*}
g\!\left(\omega\right) = \cases{\frac{1}{2 \gamma}, & $\left|\omega-\omega_0\right| \le \gamma$, \\
0, & $\left|\omega-\omega_0\right| > \gamma$.}
\label{eq:UniformDistribution}
\end{equation*}
Using \eref{eq:rC012} and \eref{eq:mucr}, we obtain
\begin{eqnarray}
C_0 = \frac{1}{\sigma^2}\frac{\left(1+\mu\sigma^2\right)\arctan\xi - \mu\sigma^2\xi}{2 \xi}, \nonumber \\
\fl C_1 = -\frac{1}{8\sigma^6}\frac{\left(1+2 \mu\sigma^2\right)\xi+\left(1+\xi^2\right)\left(\left(1+2 \mu\sigma^2\right)\arctan\frac{\xi}{2}-\left(1+3 \mu\sigma^2\right)\arctan\xi\right)}{\xi\left(1+\xi^2\right)}, \label{eq:C0mucrUniformDistribution}\\
\mu^{*} = -\frac{1}{\sigma^2} \frac{\xi + \left(\xi^2+1\right)\left(\arctan\frac{\xi}{2}-\arctan\xi\right)}{2 \xi + \left(\xi^2+1\right)\left(2 \arctan\frac{\xi}{2}-3 \arctan\xi\right)} \nonumber.
\end{eqnarray}

In the noise-free case, i.e., at $\sigma=0$, it follows from \eref{eq:a01SolutionSigma0} that
\begin{equation}
\fl r\!\left(A\right) = o\!\left(\mu\right) + \cases{A\left(\frac{\pi}{4 \gamma}-\frac{\mu}{3}\left(1-\frac{A}{\gamma}\right)\frac{\gamma-A+2\sqrt{\gamma^2-A^2}}{\gamma+\sqrt{\gamma^2-A^2}}\right), & $A\le\gamma$,\\
\frac{1}{2}\left(\sqrt{1-\frac{\gamma^2}{A^2}}+\frac{A}{\gamma}\arcsin
\left(\frac{\gamma}{A}\right)\right), & $A>\gamma$.}
\label{eq:rUniformDistribution}
\end{equation}
The critical value of the coupling is
$\varepsilon_c^{\left(1\right)} = 4\gamma\left(1+2\mu\gamma/\pi\right)/\pi + o\!\left(\mu\right)$.

\begin{figure}[h]
\centering
\includegraphics[width=17.2cm]{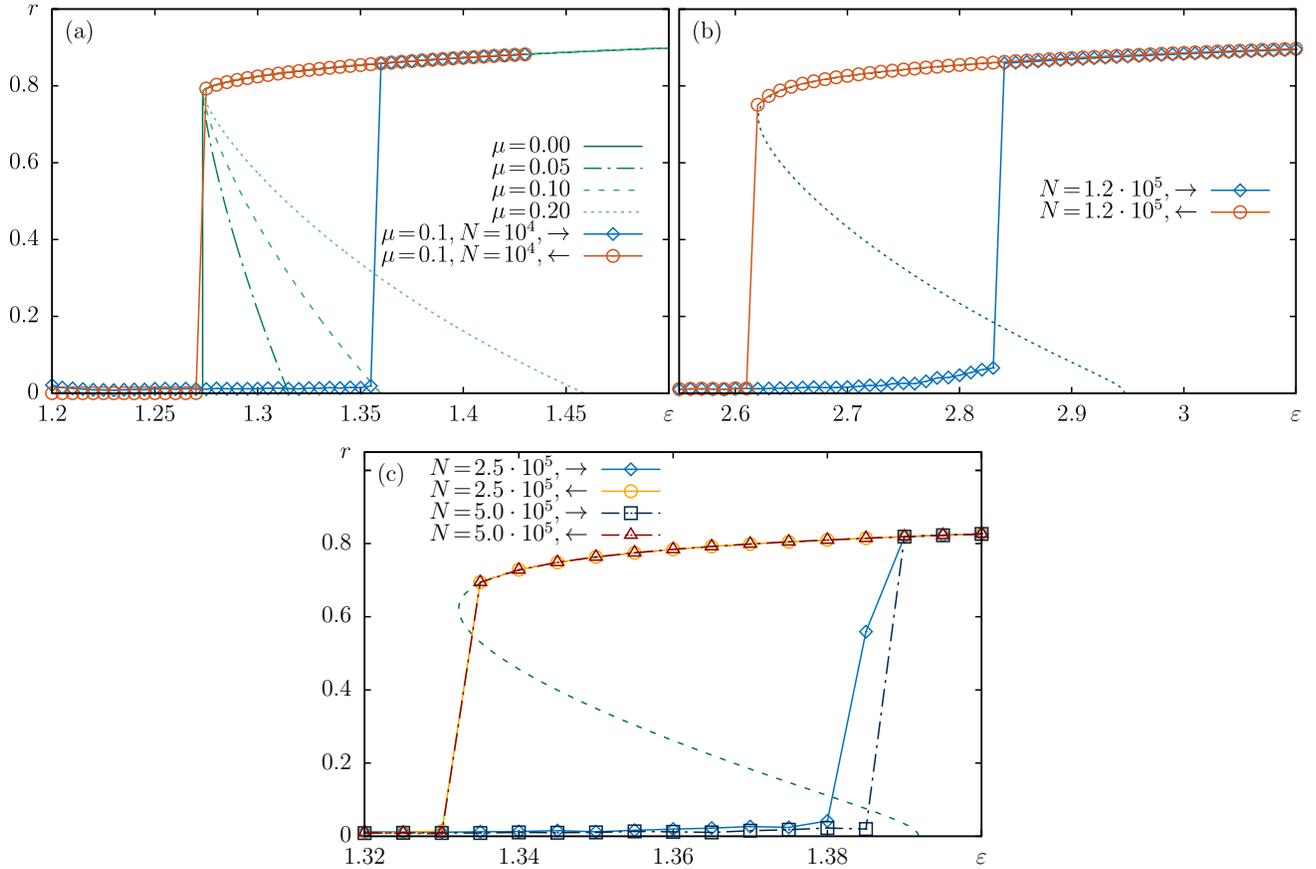}
\caption{
(a) Curves: branches of stationary solutions $r(\varepsilon)$ for different
$\mu$ in the noise-free case obtained analytically from equation \eref{eq:rUniformDistribution}.
Connected markers: branches obtained from direct numerical simulations of the ensemble of
$N=10000$ rotators with $\mu=0.1$ in a setup where the coupling parameter $\varepsilon$ gradually increases ('$\rightarrow$' label, diamond and square markers) or decreases ('$\leftarrow$' label, round and triangle markers). 
In panels (b) and (c), results of similar numerical experiments
are shown for rotators with masses $\mu=0.1$, noise amplitude $\sigma=0.2$, and two different widths
of the distribution, $\gamma=2$ in (b) and $\gamma=1$ in (c). Comparison with theoretical predictions (dashed lines)
shows particularly strong effect of finite-size fluctuations on the discontinuous transition ``asynchrony $\to$
synchrony''; the reverse transition is nearly at the point predicted by analytical theory even for not too large populations.}
	\label{fig:unif1}
\end{figure}

A remarkable feature of this distribution is that it demonstrates a discontinuous
transition without hysteresis for $\mu=\sigma=0$~\cite{Pazo-05}: at $\varepsilon=4\gamma/\pi$ the order parameter $r$
jumps from zero to $\pi/4$. Both small noise and small inertia destroy this degeneracy, although
in different ways.
For small values of
$\mu$ and vanishing noise $\sigma\to 0$, the transition is subcritical, 
see Figure~\ref{fig:unif1}(a), where 
we compare the analytical result~\eref{eq:rUniformDistribution} with direct 
numerical simulations. For small noise, in the massless case, the transition is supercritical,
but it turns to be subcritical beyond the critical mass $\mu^*$ given by~\eref{eq:C0mucrUniformDistribution}. This
situation is illustrated in Figure~\ref{fig:unif1}(b,c). Direct numerical simulations clearly show a histeresis
and regions of bistability, where both the asynchronous state and the stationary synchronous
branch are stable. Here, one can also see different effects of finite-size fluctuations on the transitions
to synchrony and back: the asynchronous state is much more sensitive to these fluctuations, which
results in a shift of the transition point to smaller values of coupling.

\section{Generalizations}\label{sec:Generalizations}
\label{sec:gen}
Above we focused on the case where the rotators differ solely by their natural frequencies
$\omega_n$, and the distribution of these frequencies is symmetric. Often, it is desirable
to consider a more general situation, where all the parameters governing the dynamics of
rotators are different (cf. a similar generalization 
of the Kuramoto model in~\cite{Vlasov-Macau-Pikovsky-14}):
\begin{equation}
\eqalign{
\mu_n \ddot{\varphi}_n + \dot{\varphi}_n &= 
\omega_n + \frac{\epsilon_n}{N} \sum_{n'=1}^{N} \sin\!\left(\varphi_{n'}-\varphi_n-\beta_n\right) + 
\sigma_n\xi_n\!\left(t\right) \\
&=\omega_n + \epsilon_n r \sin\!\left(\psi-\varphi_n-\beta_n\right) + \sigma_n\xi_n\!\left(t\right).}
\label{eq:grm}
\end{equation}
Here, we introduce two global parameters: $E$ is the average strength of coupling, and $B$ is the
characteristic phase shift in coupling,
\begin{equation*}
\epsilon_n=E+\varepsilon_n,\quad \beta_n=B+\alpha_n.
\end{equation*}
The goal is to find uniformly rotating solutions $r=\mathrm{const}$, $\dot\psi=\Omega$ in the thermodynamic limit.
For the given distribution of individual parameters $\mu_n,\omega_n,\varepsilon_n,\alpha_n,\sigma_n$,
and for values of global parameters $E$ and $B$, we look for a solution $r,\Omega$ (multiple solutions are
also possible). Similar to the consideration above, it is more convenient to fix $\Omega$ and $A=rE$, and to find
the solution in a parametric form $E=E(A,\Omega)$, $B=B(A,\Omega)$, $r=A/E(A,\Omega)$.

To accomplish this, we introduce the rotating variable $\theta=\varphi-\psi-B-\alpha$. Then \eref{eq:grm}
takes form
\begin{equation*}
\mu\ddot{\theta} + \dot{\theta} = 
\omega-\Omega - A\varepsilon \sin\!\left(\theta\right) + 
\sigma\xi\!\left(t\right).
\end{equation*}
The stationary distribution for this Langevin equation was 
analyzed in Section~\ref{sec:MatrixMethod} above, it yields the following subgroup order parameter
\begin{equation}
\fl a_{0,1}\!\left(\omega,\mu,\varepsilon,\sigma \mid A,\Omega\right) = \frac{1}{\sqrt{2 \pi}} \frac{I_{1+i\frac{\omega-\Omega}{\sigma^2}}\!
\left(\frac{A\varepsilon}{\sigma^2}\right)}
{I_{i\frac{\omega-\Omega}{\sigma^2}}\!\left(\frac{A\varepsilon}{\sigma^2}\right)}
\left(1-\mu \frac{\sigma^2}{\pi}\frac{\sin\!\left(i \pi \frac{\omega-\Omega}{\sigma^2}\right)}
{I_{-i \frac{\omega-\Omega}{\sigma^2}}\!\left(\frac{A\varepsilon}{\sigma^2}\right)
I_{i \frac{\omega-\Omega}{\sigma^2}}\!
\left(\frac{A\varepsilon}{\sigma^2}\right)}\right) \!+ o\!\left(\mu\right).
\label{eq:a01Solution1}
\end{equation}
Substituting this solution in the expression for the global order parameter $r$, we obtain
\begin{equation}
re^{-iB}=\int \dif\omega\,\dif\mu\,\dif\varepsilon\,\dif\sigma\, \dif\alpha\,
a_{0,1}^*\!\left(\omega,\mu,\varepsilon,\sigma \mid A,\Omega\right)e^{i\alpha}G(\omega,\mu,\varepsilon,\sigma,\alpha)
\label{eq:scG}
\end{equation}
where $G(\omega,\mu,\varepsilon,\sigma,\alpha)$ is a joint distribution density over the parameters of the problem.
Expression \eref{eq:scG} yields the representation of the solution in the explicit parametric form
\[
r=r(A,\Omega),\quad B=B(A,\Omega),\quad E=\frac{A}{r(A,\Omega)}.
\]
Clearly, if only some of the parameters are distributed, general expressions~\eref{eq:a01Solution1}, \eref{eq:scG}
can be simplified.

\section{Conclusion}
\label{sec:con}
In conclusion, we have developed an analytical description of stationary synchronous regimes
in a population of noisy globally coupled rotators (``oscillators with inertia''). The main analytical formula
is expression \eref{eq:a01Solution} for the subgroup order parameter of oscillators having detuning
$\nu$ to the frequency of the mean field. This expression contains only normalized detuning $\nu/\sigma^2$
and forcing $A/\sigma^2$, and is valid for small masses $\mu$. We also discussed different
limiting cases which can be straightforwardly derived from this formula. For example, for massless rotators, i.e.,
for the standard Kuramoto oscillators, we obtain an exact expression~\eref{eq:klop}.
This provides an analytic expression for stationary solutions for the Kuramoto model 
(or, more generally, for the Kuramoto--Sakaguchi model) with noise. 

Our approach is restricted to stationary solutions only, it does not capture possible regimes with
a non-constant modulus of the order parameter. The latter are essential for multimodal
distributions of
natural frequencies, in particular for the bimodal distribution considered in 
Section~\ref{sec:BiharmonicDistribution}, where periodic regimes dominate the transition to synchrony.
Another restriction of our approach is that it does not provide stability of the
nontrivial solutions:
the corresponding linearized equations have to be explored numerically (even stability analysis
of the trivial asynchronous state is rather involved, see~\cite{Acebron-Bonilla-Spigler-00}).

Finally, we have shown that the approach can be directly generalized to the case where
not only the natural frequencies of rotators are different, but also their 
masses (cf.~\cite{Komarov-Gupta-Pikovsky-14}), coupling strengths and phase shifts in coupling
(cf.~\cite{Vlasov-Macau-Pikovsky-14}), or even noise intensities. Such a generalization can be
useful for applications of mean field theory 
to random network couplings, e.g., the diversity of incoming
degrees can be modelled as a distribution of effecting coupling constants.

\ack

We thank D. Goldobin for useful discussions. Results presented in Sections~\ref{sec:Model} and \ref{sec:MatrixMethod} were supported by the RSF grant No.~17-12-01534. Results presented in Sections~\ref{sec:LimitingCases}, \ref{sec:Examples} and \ref{sec:Generalizations} were supported by the RSF grant No.~19-12-00367. Results presented in Appendix~\ref{sec:CoordinateMethod} and numerical simulations presented in Section~\ref{sec:Examples} were supported by the RFBR grant No.~19-52-12053.

\appendix

\section{Derivation of the subgroup order parameter by virtue of moment expansion}\label{sec:CoordinateMethod}

We start with a general reduction of a second-order stochastic equation
\begin{equation*}
\mu\ddot\varphi+\dot\varphi=F(\varphi,t)+\sigma\xi(t)
\label{eq:2og}
\end{equation*}
to a first-order equation, valid for small $\mu$. The probability density $P(\varphi,\dot\varphi,t)$
obeys the corresponding Fokker--Planck equation
\begin{equation}
\partial_t P+\partial_\varphi\left(\dot\varphi P\right)
+\partial_{\dot\varphi}\left(\mu^{-1}(-\dot\varphi+F)P\right)=\mu^{-2}\sigma^2\partial_{\dot{\varphi}}^2 P.
\label{eq:afpe}
\end{equation}
We employ the moment method described in~\cite{Wilemski-76}.

First, we introduce the moments in the velocity:
\begin{equation*}
w_m\!\left(\varphi,t\right) = 
\int_{-\infty}^{\infty}\dif\dot{\varphi} \dot{\varphi}^m P\!\left(\varphi,\dot{\varphi},t\right).
\label{eq:Momentsdphi}
\end{equation*}
According to~\eref{eq:afpe}, these moments obey following equations:
\begin{eqnarray}
\partial_t w_0 + \partial_{\varphi} w_1 &= 0, \label{eq:MomentsdphiSystem1}\\
w_1 + \mu\partial_t w_1 &= F w_0 - \mu\partial_{\varphi}w_2, \label{eq:MomentsdphiSystem2}\\
w_m + \frac{\mu}{m}\partial_t w_m &= F w_{m-1} - \frac{\mu}{m}\partial_{\varphi}w_{m+1} + 
\left(m-1\right)\frac{\sigma^2}{\mu}w_{m-2} \mbox{ for }m\ge 2. \label{eq:MomentsdphiSystem3}
\end{eqnarray}
In order to employ the smallness of parameter $\mu$,
it is convenient to introduce rescaled moments
\begin{equation*}
w_m = \cases{\frac{1}{\mu^{m/2}}W_m & for even $m$, \\
\frac{1}{\mu^{\left(m-1\right)/2}}W_m & for odd $m$.}
\label{eq:NormalizedMomentsdhi}
\end{equation*}
In terms of moments $W_m$, equations \eref{eq:MomentsdphiSystem1}--\eref{eq:MomentsdphiSystem3}
can be rewritten in a form free of singularities
\begin{eqnarray*}
\partial_t W_0 + \partial_{\varphi} W_1 = 0, \\
W_1 = F W_0 - \partial_{\varphi} W_2 - \mu\partial_t W_1, \\
\fl W_m = \left(m-1\right)\sigma^2 W_{m-2} + 
\mu\left(F W_{m-1} - \frac{1}{m}\partial_{\varphi}W_{m+1} - \frac{1}{m}\partial_t W_m\right) &
\mbox{ for even } m, \\
\fl W_m = \left(m-1\right)\sigma^2 W_{m-2} + F W_{m-1} - 
\frac{1}{m}\partial_{\varphi}W_{m+1} - \frac{\mu}{m}\partial_t W_m &\mbox{ for odd } m.
\label{eq:NormalizedMomentsdphiSystem}
\end{eqnarray*}
As we are looking for the first order corrections in $\mu$, we rewrite this system
keeping the relevant terms only
\begin{eqnarray}
\partial_t W_0 + \partial_{\varphi} W_1 = 0, \nonumber \\
W_1 = F W_0 - \partial_{\varphi} W_2 - \mu\partial_t W_1, \nonumber \\
W_2 = \sigma^2 W_0 + \mu\left(F W_1 - \frac{1}{2}\partial_{\varphi}W_3 - \frac{1}{2}\partial_t W_2\right), \nonumber\label{eq:NormalizedMomentsdphiSystem4}\\
W_3 = 2\sigma^2 W_1 + F W_2 - \frac{1}{3}\partial_{\varphi}W_4 + O\!\left(\mu\right), \nonumber \\
W_4 = 3\sigma^2 W_2 + O\!\left(\mu\right) \nonumber.
\end{eqnarray}
Now, starting with substituting the expression for $W_4$ in the equation for $W_3$,
we find
\begin{eqnarray}
W_3 = 2\sigma^2 W_1 + F W_2 - \sigma^2\partial_{\varphi}W_2 + O\!\left(\mu\right), \nonumber\\
\fl W_2 = \sigma^2 W_0 + \mu\left[-\frac{\sigma^2}{2}\partial_t W_0 + 
F\left(F W_0 - \sigma^2\partial_{\varphi}W_0\right) - \frac{\sigma^2}{2}\partial_{\varphi}\left(F W_0\right) \right. \nonumber\\
\left. + \frac{\sigma^4}{2}\partial_{\varphi}^2 W_0 - \sigma^2\partial_{\varphi}\left(F W_0 - 
\sigma^2\partial_{\varphi}W_0\right)\right] + O\!\left(\mu^2\right), \nonumber\\
\fl W_1 = F W_0 - \sigma^2\partial_{\varphi}W_0 + \mu\left[-\left(\partial_t F + 
F\partial_{\varphi}F\right)W_0 + 
\sigma^2\left(\partial_{\varphi}F\right)\partial_{\varphi}W_0\right] + O\!\left(\mu^2\right).\nonumber
\label{eq:ReducedFokkerPlanckDer}
\end{eqnarray}
Finally, in the first order in $\mu$, we obtain the following Fokker--Planck equation for the
distribution density of the phases $\rho(\varphi,t)\equiv W_0(\varphi,t)$:
\begin{equation*}
\partial_t \rho + \partial_{\varphi}\left[\left(F\left(1-\mu\partial_{\varphi}F\right)-
\mu\partial_t F\right)\rho\right]= \sigma^2\partial_{\varphi}
\left[\left(1-\mu\partial_{\varphi}F\right)\partial_{\varphi}\rho\right].
\label{eq:ReducedFokkerPlanck}
\end{equation*}
The corresponding Langevin equation reads
\begin{equation*}
\dot{\varphi} = F - \mu\left(\partial_t + F\partial_{\varphi} + 
\frac{\sigma^2}{2}\partial_{\varphi}^2\right)
F + \sigma\sqrt{1 - \mu\partial_{\varphi}F}\,\xi\!\left(t\right).
\label{eq:EffectiveLangevinEquation}
\end{equation*}

Next, similarly to the procedure described in Section~\ref{sec:Model},
we transform to the rotating reference frame by introducing $\theta=\varphi-\psi$, where $\dot\psi=\Omega$.
In this reference frame we set $F=\nu-A\sin\theta$ and look for a stationary solution $\rho_0(\theta)$, which satisfies
the following equation
\begin{equation}
0 = -\partial_{\theta}\left[\left(1+\mu A\cos\theta\right)\left(\nu-A\sin\theta\right)\rho_0\right] 
+ \sigma^2 \partial_{\theta}\left[\left(1+\mu A\cos\theta\right)\partial_{\theta}\rho_0\right].
\label{eq:ReducedStationarySystem}
\end{equation}

Solution of~\eref{eq:ReducedStationarySystem} satisfying periodicity 
$\rho_0\!\left(\theta\right)=\rho_0\!\left(\theta+2\pi\right)$, 
reads (cf~\cite{Stratonovich-63})
\begin{equation}
\rho_0\!\left(\theta \mid \omega\right)=\frac{1}{Z}e^{V\!\left(\theta \right)/\sigma^2}
\int_{\theta}^{\theta+2\pi}\dif\theta'\left(1-\mu A\cos\theta'\right)
e^{-V\!\left(\theta'\right)/\sigma^2} + o\!\left(\mu\right),
\label{eq:ReducedStationarySolution}
\end{equation}
where $V\!\left(\theta\right)=\nu\theta+A\cos\theta$, 
and $Z$ is a normalization constant,
\begin{eqnarray*}
\fl Z=2\pi^2 e^{-\pi\nu/\sigma^2}\left\{2 I_{i\frac{\nu}{\sigma^2}}\!\left(-\frac{A}{\sigma^2}\right) I_{-i\frac{\nu}{\sigma^2}}\!\left(-\frac{A}{\sigma^2}\right)\right.\nonumber\\
\fl \left. {}-\mu A\left[I_{1+i\frac{\nu}{\sigma^2}}\!\left(-\frac{A}{\sigma^2}\right) I_{-i\frac{\nu}{\sigma^2}}\!\left(-\frac{A}{\sigma^2}\right)+I_{i\frac{\nu}{\sigma^2}}\!\left(-\frac{A} {\sigma^2}\right) I_{1-i\frac{\nu}{\sigma^2}}\!\left(-\frac{A}{\sigma^2}\right)\right]\right\}+o\!\left(\mu\right).
\label{eq:W1Normalization}
\end{eqnarray*}
(see formulas 6.681.3 and 8.511.4 in \cite{Gradshteyn-Ryzhik-07}). 
Similar integrals appear for the order parameter $\langle e^{i\theta}\rangle$ 
by virtue of
\eref{eq:ReducedStationarySolution}; the resulting expression 
coincides with that for $a_{0,1}\!\left(\omega\right)$ (equation~\eref{eq:a01Solution})
up to normalization.

~\par~\par

%\bibliographystyle{iopart-num}
%\bibliography{biblio_stgcr}
\providecommand{\newblock}{}

\end{document}